\documentclass[aps,prc,preprint,floatfix,showkeys,showpacs]{revtex4-1}

\usepackage{amsmath}
\usepackage{graphicx}
\DeclareGraphicsExtensions{.ps}

\def\gtorder{\mathrel{\raise.3ex\hbox{$>$}\mkern-14mu
	\lower0.6ex\hbox{$\sim$}}}
\def\ltorder{\mathrel{\raise.3ex\hbox{$<$}\mkern-14mu
	\lower0.6ex\hbox{$\sim$}}}

\begin{document}

\title{Target mass corrections to parity-violating DIS}



\author{T. Hobbs}
\affiliation{Department of Physics and Center for Exploration of Energy
and Matter, Indiana University, Bloomington, IN 47405, USA}

\begin{abstract}
We examine the implications of several parameterizations of so-called
target mass corrections (TMCs) for the physics of parity-violating
deeply inelastic scattering (DIS), especially at high values of the
momentum fraction $x$. We consider the role played by perturbative
corrections in $\alpha_S$ in modifying TMCs; we explicitly
calculate these corrections at both the level of the individual
electroweak structure function (SF), as well as in the observables
of parity-violating DIS. TMCs augment an inventory of previously
studied corrections that become sizable at low $Q^2$ (finite-$Q^2$
corrections), and we give special attention to the effects that might
lead to the violation of the approximate equality
$R^{\gamma Z} = R^{\gamma}$.
\end{abstract}

\maketitle

\section{Introduction}

Recent developments in the drive to uncover the details of hadronic structure
via high energy electron scattering have increased the interest in performing
more precision measurements of nucleonic observables. Among such efforts is
the push to extract partonic information at large values of the scaling
variable Bjorken $x$ \cite{Souder:2005tz}. Unfortunately, it is often at high
$x$ that various corrections that scale as $1/Q^2$ or log($Q^2$) become
significant and render measurements and their interpretation problematic.
Typical of these kinds of corrections are target mass corrections (TMCs),
which must be implemented as one moves away from the high-energy Bjorken limit
at which the mass of the target nucleon may be neglected. The role of TMCs in
electron-nucleon deeply inelastic scattering (DIS) at next-to-leading order
(NLO) in $\alpha_S$ is investigated here.

\section{TMC prescriptions}
\begin{figure}[t]
\includegraphics[height=7.5cm,angle=270]{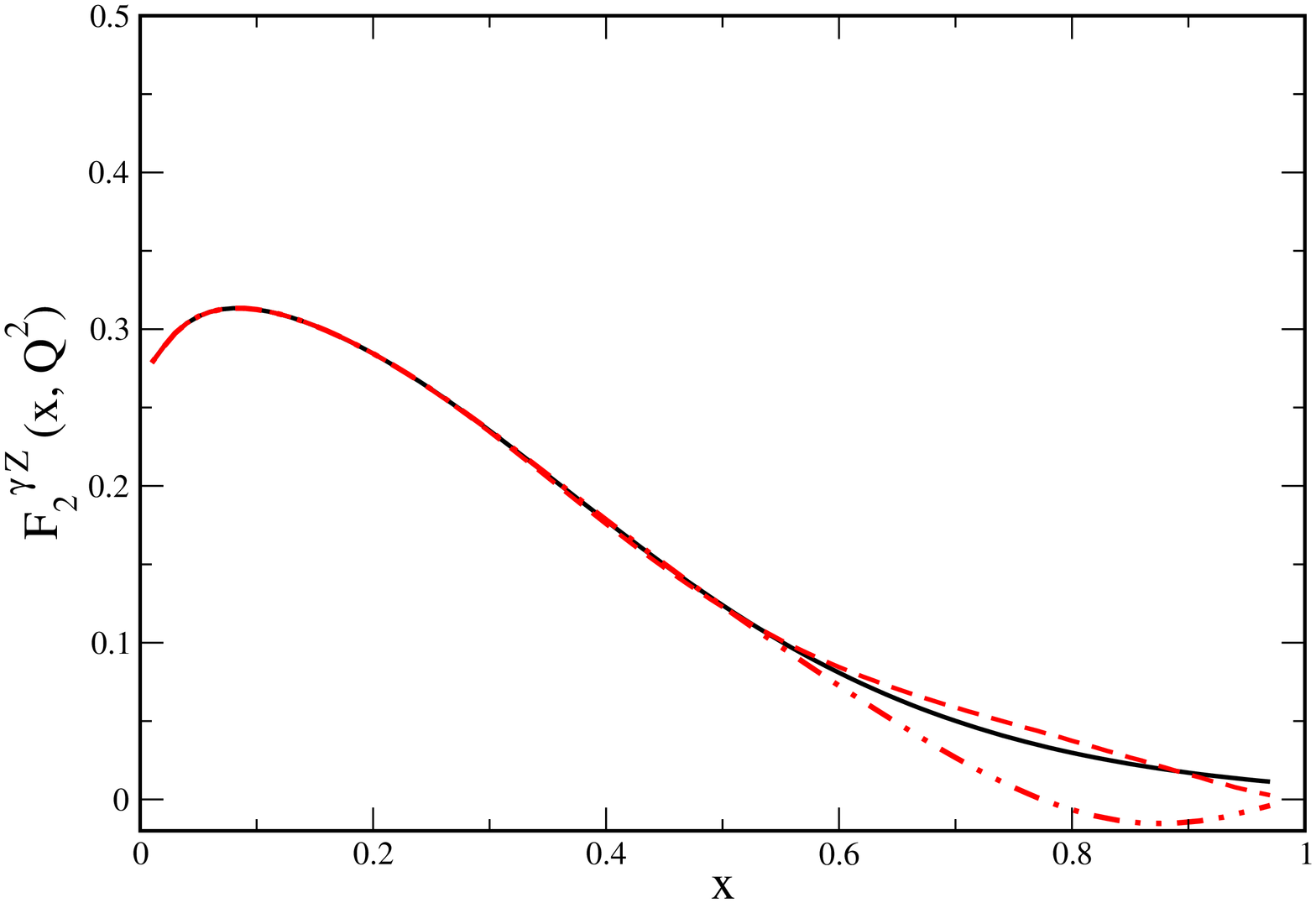}
\includegraphics[height=7.5cm,angle=270]{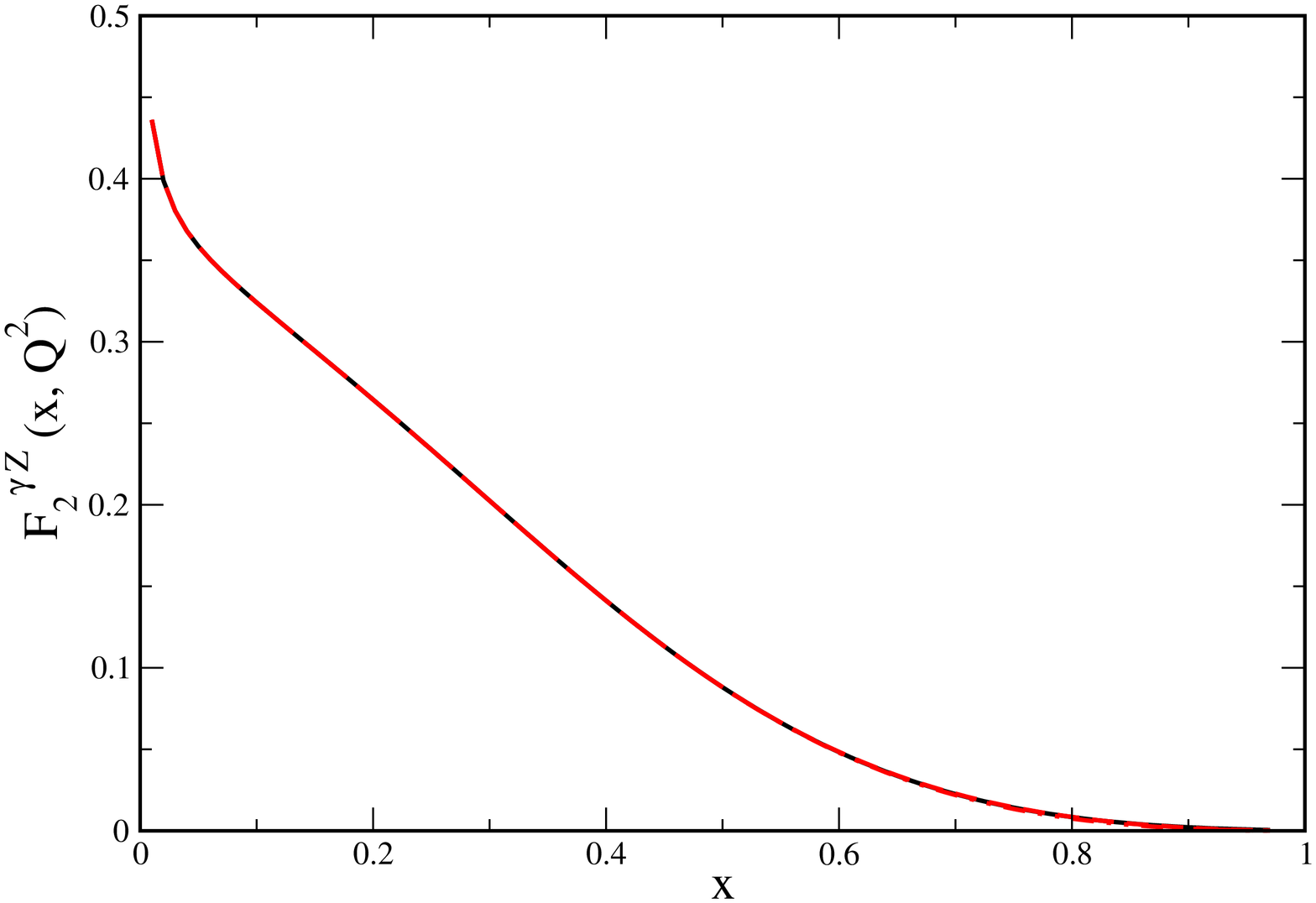}
\caption{We plot $F_2^{\gamma Z}$ over the full kinematic range
(i.e., $x \in [0, 1]$) for fixed $Q^2$ following the implementation
of the OPE (solid), O($1/Q^2$) (dashed) and O($1/Q^4$) (dot-dashed)
corrections. The non-physical threshold behavior and negative down-turn
are evident in the OPE and O($1/Q^4$) curves, respectively.}
\label{fig:Fig_1}
\end{figure}

We calculate the TMCs in three main prescriptions: the conventional,
leading-twist (LT) operator product expansion (OPE) of Georgi/Politzer, 
1/$Q^2$ expansions of these LT corrections to various orders (which we take 
to be one formal prescription), and collinear factorization (CF).
Historically well-established, the OPE has been thoroughly deployed
in the careful analysis of TMCs to the (un)polarized structure functions (SFs)
of DIS at leading perturbative order and at twist-2 and twist-3
\cite{Blumlein:1998nv}. In spite of this, there is a particular shortfall of
the LT OPE treatment to provide the proper behavior of the mass-corrected SFs
at kinematic threshold. This can be made evident by considering the corrected 
electroweak SFs as calculated in the LT OPE, 
and comparing these with the perturbative expansions of the same to 
O(1/$Q^2$) and O(1/$Q^4$). As an example of one such SF that enters
into the parity-violating asymmetry $A^{PV}$, $F_2^{\gamma Z}$ is
calculated at NLO in $\alpha_S$ at two scales: $Q^2 =$ 2,
10 GeV$^2$. For simplicity, the left (right) column in all following
plots corresponds to $Q^2 =$ 2 (10) GeV$^2$. Here and in the following
calculations, input SFs (that is, before the TMC is applied) are 
calculated according to standard quark-parton model (QPM) definitions with 
electroweak charges as given in the PDG. We notice that expansions to
both orders of $1/Q^2$ diverge from the LT calculation most at lowest 
$Q^2$ and highest $x$ as one might expect. This effect is illustrated
in Figure \ref{fig:Fig_1}.

To understand this behavior we recall that the corrected SFs in the
Georgi/Politzer formalism are extracted via inverse Mellin transforms of
the LT expansions of individual SF moments \cite{Georgi:1976ve,
Nachtmann:1973mr} to obtain for spin-unpolarized DIS
\begin{eqnarray}
F_1^{TMC}(x) = \frac{x}{\xi \rho} F_1^{(0)}(\xi) + \frac{\mu x^2}{\rho^2}
 \int^1_{\xi} du \frac{F_2^{(0)}(u)}{u^2} +
\frac{2 \mu^2 x^3}{\rho^3} \int^1_{\xi} dv (v - \xi) \frac{F_2^{(0)}(v)}{v^2}
\nonumber \\
F_2^{TMC}(x) = \frac{x^2}{\xi^2 \rho^3} F_2^{(0)}(\xi) + \frac{6 \mu x^3}{\rho^4}
 \int^1_{\xi} du \frac{F_2^{(0)}(u)}{u^2} +
\frac{12 \mu^2 x^4}{\rho^5} \int^1_{\xi} dv (v - \xi) \frac{F_2^{(0)}(v)}{v^2}
\nonumber \\
F_3^{TMC}(x) = \frac{x}{\xi \rho^2} F_3^{(0)}(\xi) + \frac{2 \mu x^2}{\rho^3}
 \int^1_{\xi} du \frac{F_3^{(0)}(u)}{u},
\label{eq:GP}
\end{eqnarray}
where $\mu = M^2 / Q^2$, and $\rho = \sqrt{1 + 4 x^2 M^2 / Q^2}$; we
note also that the SFs appearing in Eqn (\ref{eq:GP}) also contain
some omitted dependence on $Q^2$. The LT results for $F_2^{\gamma Z}$ are
plotted as the solid curves of Figure \ref{fig:Fig_1}. The associated dashed/dot-dashed
curves are the O($1/Q^2$) and O($1/Q^4$) expansions of the LT result,
respectively; we discuss the details of this calculation below.

A troubling ailment of the standard TMC treatment is evident in Figure 
\ref{fig:Fig_1}: as one approaches the lowest accessible values of $Q^2$,
the mass-corrected SFs attain zero only in the non-physical
region $x > 1$; that is, momentum conservation considerations would dictate that the
electroweak SFs equal zero for $x \geq 1$. That this does not happen
is an indication that the OPE incorporates some non-physical behavior in the
move to low $Q^2$. One can understand the origin of this non-physical behavior
by recognizing that the standard GP prescription rescales the PDFs
appearing in the QPM expressions for the SFs via the parameter
\begin{equation}
\xi (x, Q^2) = -\frac{q^+}{p^+} = \frac{2x}{1 + \rho},
\end{equation}
for which we note $\xi (x, Q^2) \leq x$ for $\mu
\geq 0$. As the SFs are monotonically decreasing functions
of $x$ for a given fixed $Q^2$, the rescaling forced by this 
procedure dictates that the corrected SFs are upward-shifted
in magnitude, attaining zero only for $\xi (x, Q^2) = 1$. This
occurs only for non-physical $x \geq 1$.

In the literature this issue has come to be referred to as
the `threshold problem' \cite{Schienbein:2007gr}. Several alternative
prescriptions have been proposed to ameliorate this issue, including
so-called collinear factorization, as well as a less formal approach
that depends upon an O(1/$Q^2$) expansion of the LT OPE. The latter
is motivated by the logic that it is the omission of the contributions
from non-leading twist in the OPE that induces the threshold problem.
Then, supposing that the O(1/$Q^2$) expansions simulate the higher
twist contribution, one might anticipate that such an expansion
would cure the non-physical behavior at large $x$. Again, a
comparison of these perturbative expansions with the LT OPE are given
in Figure \ref{fig:Fig_1} in the case of $F_2^{\gamma Z}$.

\begin{figure}[t]
\includegraphics[height=7.5cm,angle=270]{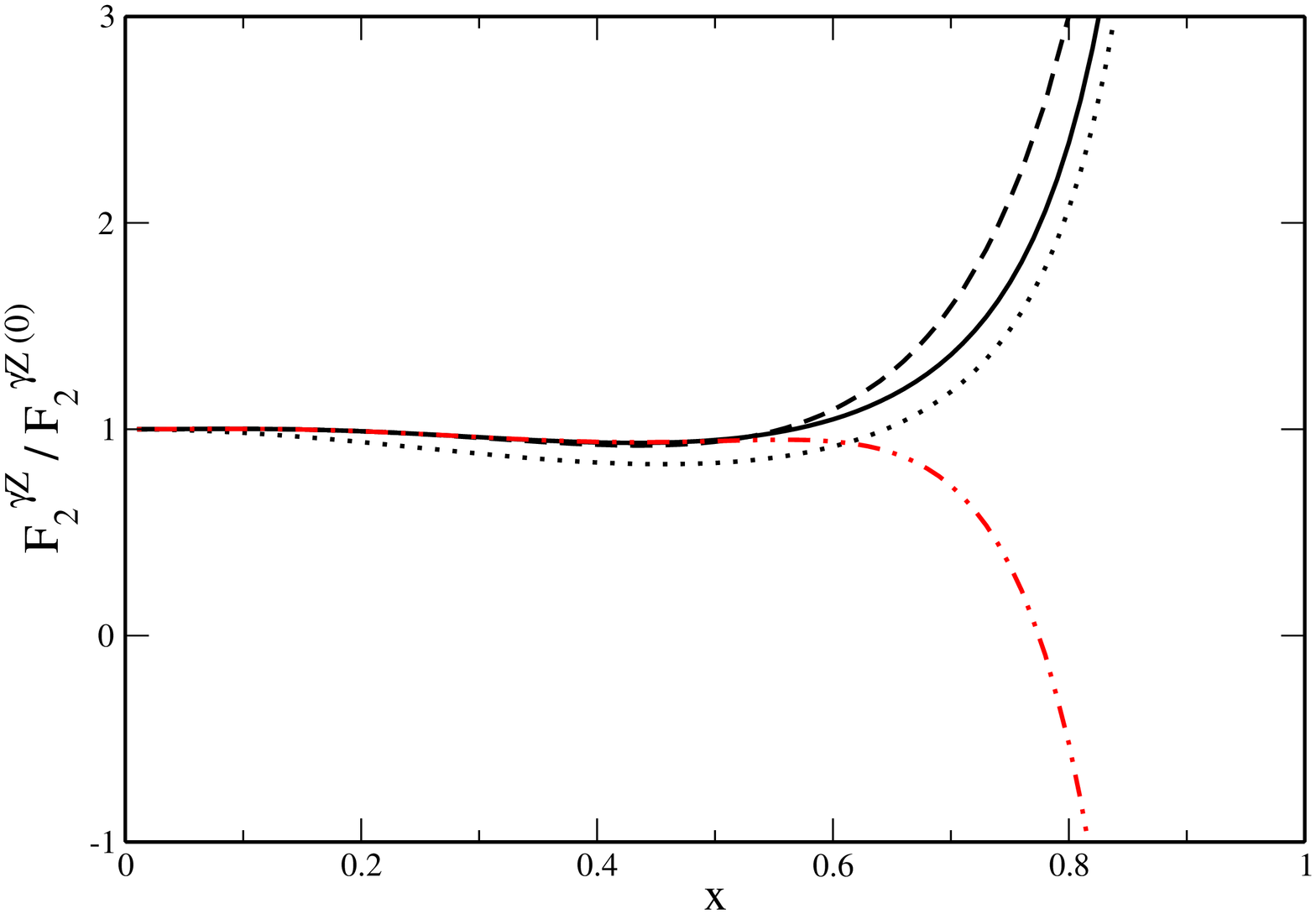}
\includegraphics[height=7.5cm,angle=270]{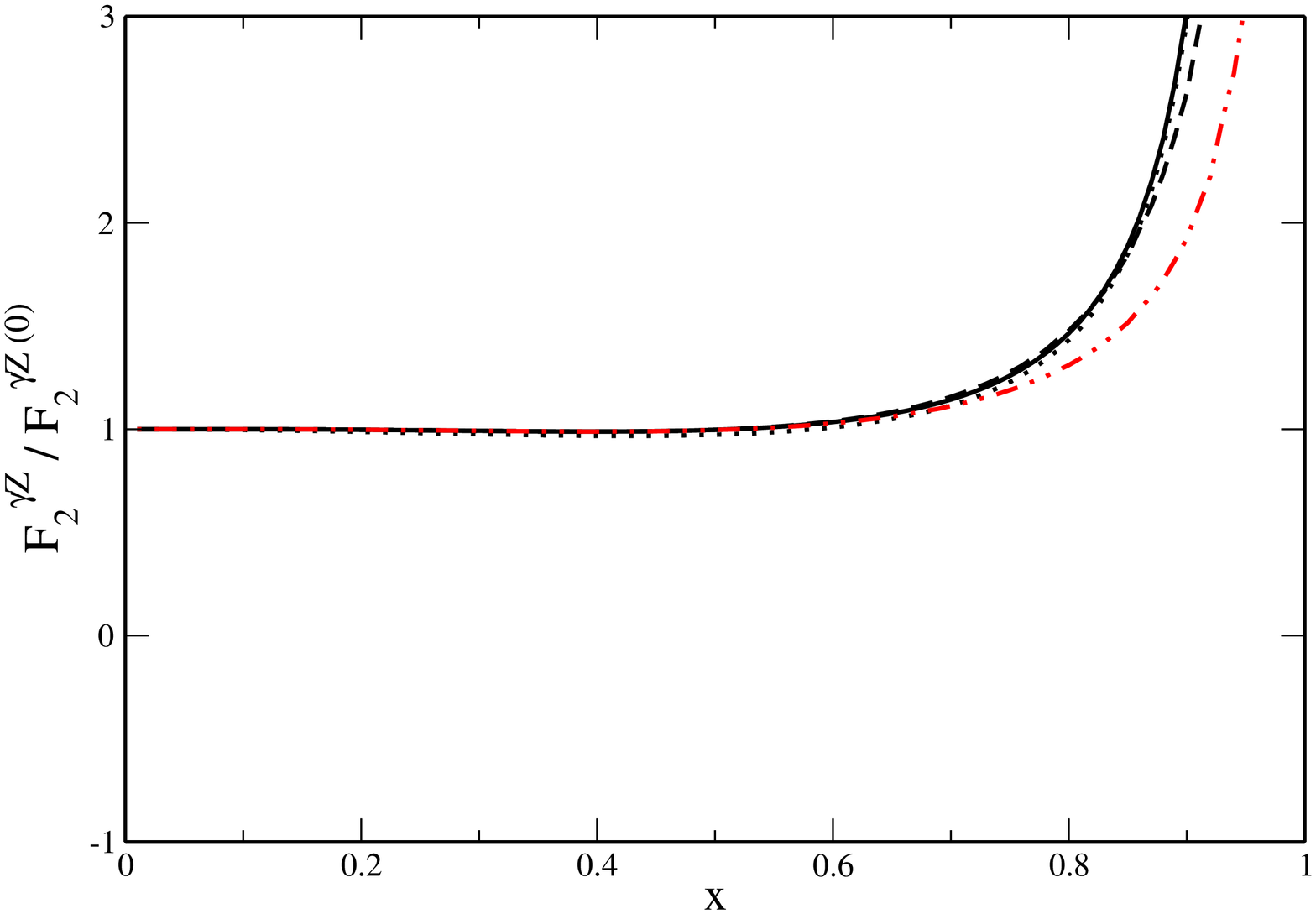}
\caption{When plotted as a ratio $F^{TMC} / F^{(0)}$, the negative downturn at
low $Q^2$ obtained in the $O(1/Q^4)$ (dot-dashed) expansion is more evident. We compare
this to the OPE (solid), O($1/Q^2$) (dashed), and CF (dotted) prescriptions.}
\label{fig:Fig_2}
\end{figure}

Via this treatment, we may expand Eqn (\ref{eq:GP}) to LO in $1/Q^2$ to
obtain \cite{Kulagin:2004ie}
\begin{eqnarray}
F_1^{TMC}(x,Q^2) = (1 - \mu x^2) F_1^{(0)}(x)
+ \mu x^3 \cdot (\frac{2}{x} \int_x^1 \frac{dz}{z^2} F_1^{(0)}(z) 
- \frac{\partial}{\partial x} F_1^{(0)}(x)) \nonumber \\ 
F_2^{TMC}(x,Q^2) = (1 - 4 \mu x^2) F_2^{(0)}(x)
+ \mu x^3 \cdot (6 \int_x^1 \frac{dz}{z^2}
F_2^{(0)}(z) - \frac{\partial}{\partial x} F_2^{(0)}(x)) \nonumber \\ 
F_3^{TMC}(x,Q^2) = (1 - 3 \mu x^2) F_3^{(0)}(x) +
 \mu x^3  \cdot (\frac{2}{x} \int_x^1 \frac{dz}{z^2}
F_3^{(0)}(z) - \frac{\partial}{\partial x} F_3^{(0)}(x)),
\label{eq:OQ2}
\end{eqnarray}
where there is again an understood $Q^2$ dependence for all SFs
on the right. The result of calculating $F_2^{\gamma Z}$ as
given in Eqn (\ref{eq:OQ2}) over a range of $x$ for low and intermediate
$Q^2$ is given as the dashed curves in Figure \ref{fig:Fig_1}. While
there has been interest expressed in the literature for attempting to
solve the non-physical behavior at kinematic threshold with such an
expansion, this procedure has only previously been carried out to
O($1/Q^2$). It is therefore sensible to carry out the expansion to an
additional perturbative order in $1/Q^2$ to gauge if this provides
further improvement in forcing the corrected SFs to observe the limit
$F_i^{TMC} (x = 1, Q^2) = 0$. Expanding the expressions in Eqn
(\ref{eq:GP}) still further, we get
\begin{eqnarray}
F_1^{O(1/Q^4)} (x, Q^2) = \dots \mu^2 (2 x^3 g_2 (x, Q^2)
     - 4 x^4 h_2 (x, Q^2) + x^3 F_2^{(0)} (x, Q^2) \nonumber \\
     + 3 x^4 F_1^{(0)} (x, Q^2) + 3 x^5 F_1^{'(0)} (x, Q^2)
     + \frac{x^6}{2} F_1^{''(0)} (x,Q^2)) \nonumber \\
\label{eq:OQ4_F1}
\end{eqnarray}
\begin{eqnarray}
F_2^{O(1/Q^4)} (x, Q^2) = \dots \mu^2 (12 x^4 g_2 (x, Q^2)
     - 48 x^5 h_2 (x, Q^2) + 23 x^4 F_2^{(0)} (x, Q^2) \nonumber \\
     + 6 x^5 F_2^{'(0)} (x, Q^2) + \frac{x^6}{2} F_2^{''(0)} (x,Q^2))
\label{eq:OQ4_F2}
\end{eqnarray}
\begin{eqnarray}
F_3^{O(1/Q^4)} (x, Q^2) = \dots \mu^2 (-12 x^4 h_3 (x, Q^2)
     + 13 x^4 F_3^{(0)} (x, Q^2) + 5 x^5 F_3^{'(0)} (x, Q^2) \nonumber \\
     + \frac{x^6}{2} F_3^{''(0)} (x,Q^2)).
\label{eq:OQ4_F3}
\end{eqnarray}
The dot-dashed curves of Figure \ref{fig:Fig_1} are obtained then from plotting
$F_2^{\gamma Z}$ from Eqn (\ref{eq:OQ4_F2}) similarly to the
GP and O($1/Q^2$) curves. Notable here is the fact that while the O($1/Q^4$) expansion
predictably forces the corrected SFs more strongly to observe the zero limit at the
kinematic threshold, it introduces a separate, non-physical behavior distinct from
the threshold problem. This becomes obvious at lowest $Q^2$ and highest $x$
and is thus an artifact of the regime in which the expansion parameter
$1/Q^2$ becomes large. In Figure \ref{fig:Fig_1}, the corrected SF becomes
negative for $x \geq 0.8$. We illustrate this effect more explicitly in Figure
\ref{fig:Fig_2}, in which we plot a ratio of corrected to uncorrected SFs.
As the cross section is expressed in terms of these electroweak SFs,
and is by the optical theorem a non-negative physical observable, we demand
that the SFs be positive-definite as well. By this logic, the negative downturn evident
in Figures \ref{fig:Fig_1}, \ref{fig:Fig_2}, etc., is prohibited. That the higher
order terms of the $1/Q^2$ expansion induce this non-physical behavior calls into
question the unlimited validity of such an approach -- at least when extended to too
low $Q^2$.

For completeness, we calculate effects in physical observables using the previously
mentioned CF approach also. This method relies upon a choice of frame such that 
(in DIS) the virtual photon and nucleon light-cone momenta are collinear;
the resulting cross section and SFs may then be expressed as a convolution
of a perturbatively calculable hard coefficient with the universal PDFs in
a fashion that directly incorporates the nucleon mass \cite{Accardi:2008ne,
Accardi:2009md}:
\begin{equation}
F_i (x, Q^2) = \sum_f \int_{x_{min}}^{x_{max}} \frac{dx}{x} h_i^f (x_f, Q^2)
\phi_{f/N} (x, Q^2).
\end{equation}
Here the bounds of integration $x_{min}, x_{max}$ are fixed by the
mass-dependent kinematics. In the case of the SF $F_2$, this procedure yields
a correction of the form
\begin{equation}
F_2 (x, Q^2) = \frac{x}{x_f} \frac{\rho_f^2}{\rho^2} h_2^f \otimes
\phi_{f/N} (\xi),
\end{equation}
in which $x_f, \rho_f$ are partonic analogs of Bjorken x and the previously 
defined $\rho$ parameter.

\section{TMCs in the observables of DIS}
Ultimately, we wish to obtain a prescription-independent estimate of the magnitude
of the effect of TMCs in measurements of typical DIS observables, such as the
parity-violating asymmetry particular to electron-proton scattering. As far as
experimental efforts to determine the high-$x$ structure of the nucleon are
concerned, there is no {\it a priori} knowledge of the size of the possible
contribution from TMCs, and a thorough theoretical analysis is needed to ensure
that the extraction of desired signals (such as the density function ratio
$d/u$) is not imperiled.

As noted, there is considerable experimental interest in more precise determinations
of the sensitivity of the parity-violating asymmetry $A^{PV}$ to the flavor
structure of the nucleon. For electromagnetic and interference currents, the
asymmetry may be expressed as \cite{Hobbs:2008mm}
\begin{equation}
A^{\rm PV}
= - \left( {G_F Q^2 \over 4 \sqrt{2} \pi \alpha} \right)
  \left[ g^e_A\ Y_1\ \frac{F_1^{\gamma Z}}{F_1^{\gamma}}\
     +\ {g^e_V \over 2}\ Y_3\ \frac{F_3^{\gamma Z}}{F_1^{\gamma}} 
  \right]\ ,
\label{eq:APV}
\end{equation}
in which the parameters $Y_1, Y_3$ are given by
\begin{eqnarray}
\label{eq:Y}
Y_1
= \frac{ 1+(1-y)^2-y^2 (1-r^2/(1+R^{\gamma Z})) - 2xyM/E }
        { 1+(1-y)^2-y^2 (1-r^2/(1+R^\gamma)) - 2xyM/E } 
   \left( \frac{1+R^{\gamma Z}}{1+R^\gamma} \right)\ ,  \nonumber \\
Y_3
= \frac{ 1-(1-y)^2 }
        { 1+(1-y)^2-y^2 (1-r^2/(1+R^\gamma)) - 2xyM/E }
   \left( \frac{r^2}{1+R^\gamma} \right)\ .
\end{eqnarray}
In particular we note that $Y_1, Y_3$ depend upon the ratios of
longitudinal/transverse virtual photon cross sections, 
\begin{equation}
R^{\gamma (\gamma Z)}\
\equiv\ \frac{\sigma_L^{\gamma (\gamma Z)}}{\sigma_T^{\gamma (\gamma Z)}}\ 
=\ r^2 \frac{F_2^{\gamma (\gamma Z)}}{2x F_1^{\gamma (\gamma Z)}} - 1\ .
\label{eq:Rdef}
\end{equation}
Consequently, to understand the mass effect in the full asymmetry,
it is important first to grasp TMCs in the setting of the
electroweak $R$ parameters.

There is a standing corpus of phenomenology \cite{Kulagin:2007ju,
Whitlow:1990gk} to describe the electromagnetic ratio $R^{\gamma}$,
but the corresponding interference quantity $R^{\gamma Z}$ is largely
undetermined. In the interest of controlling uncertainties induced by
$R^{\gamma Z} \neq R^{\gamma}$, an understanding of the physics that
might differently break the partonic Callan-Gross relations for purely
electromagnetic and interference processes is needed.
Insofar as Callan-Gross is strictly observed by the parton model at LO,
we expect $R^{\gamma Z} = R^{\gamma}$ to be broken by perturbative corrections
in $\alpha_S$, possible contributions from beyond twist-4, TMCs, and other
non-perturbative physics. In an effort to understand the role of NLO corrections
and TMCs in producing $R^{\gamma Z} \neq R^{\gamma}$, we plot a ratio of the
interference to electromagnetic $R$ parameters in the presence of mass corrections
and at NLO.
\begin{figure}[t]
\includegraphics[height=7.5cm,angle=270]{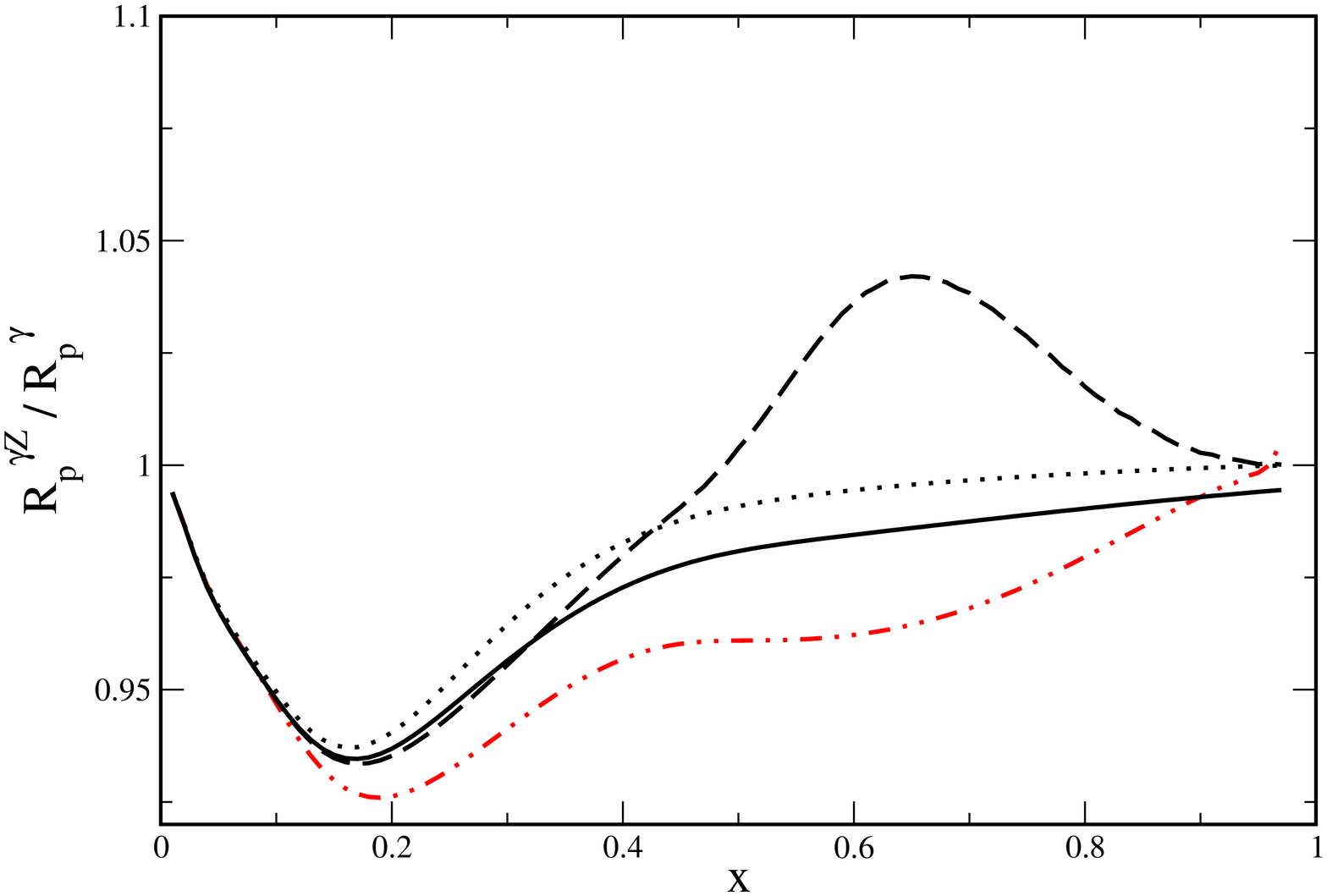}
\includegraphics[height=7.5cm,angle=270]{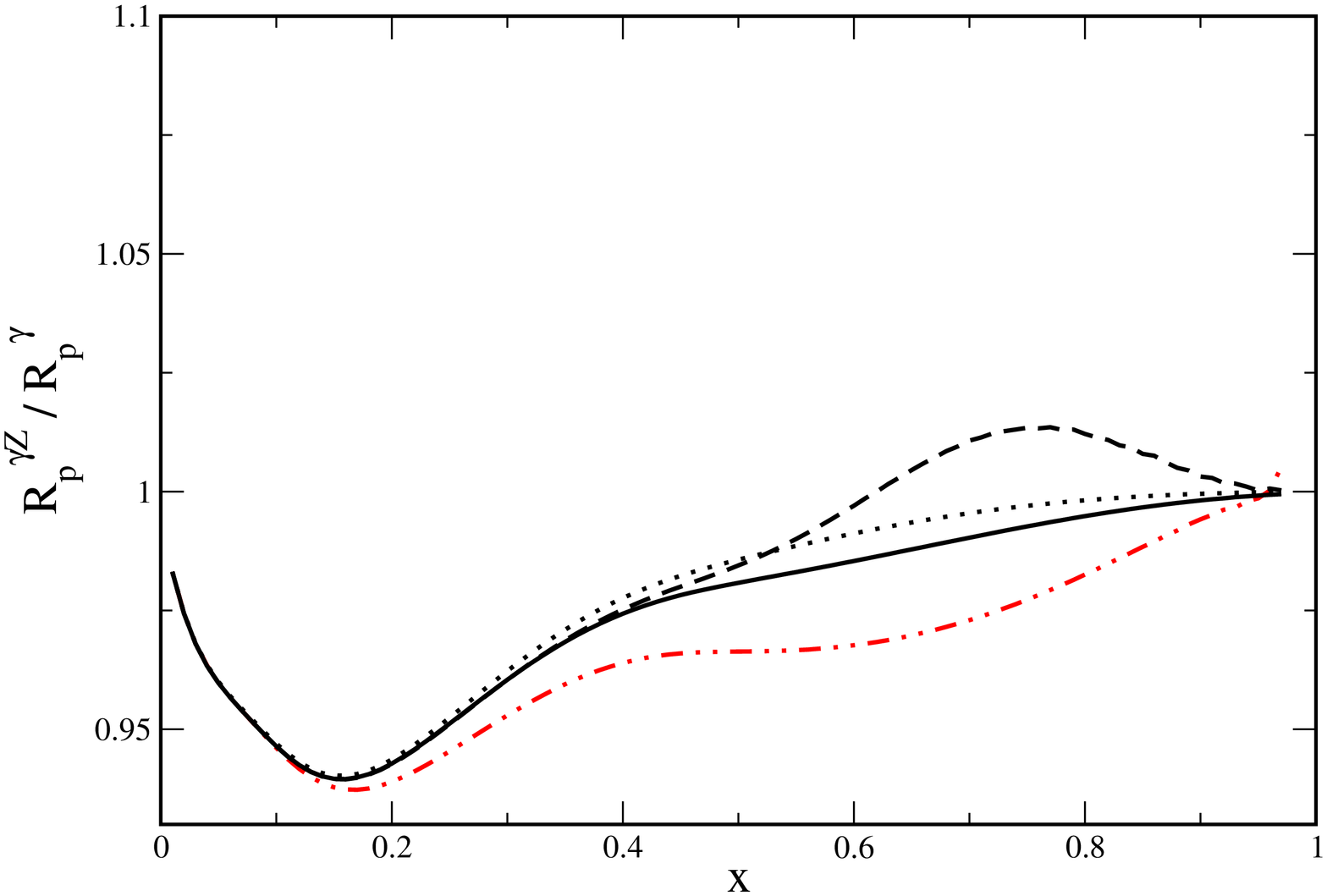}
\caption{An illustration of the breaking of the $R^{\gamma Z} = R^{\gamma}$ induced
by the target mass and perturbative corrections. We plot a ratio $R^{\gamma Z} / R^{\gamma}$ 
as obtained in proton scattering. The dot-dashed curve is generated in the absence of TMCs 
but at NLO, whereas the solid (OPE), dashed ($1/Q^2$ expansion), and dotted (CF) curves
include the nucleon mass effect as well.}
\label{fig:Fig_3}
\end{figure}
We compute this ratio using the definitions of Eqn (\ref{eq:Rdef}),
in which the parton model expressions for the SFs are typical of
proton (Figure \ref{fig:Fig_3}) and deuteron (Figure \ref{fig:Fig_4}) scattering.
The results, which are largely independent of $Q^2$, establish the largest
deviation from $R^{\gamma Z} / R^{\gamma} = 1$ at low $x$, where the effect
has a magnitude  $\approx 5 \%$. This becomes more shallow as one moves to
highest $x$; also of note is the fact that TMCs undermine the correction to
$R^{\gamma Z} / R^{\gamma}$ due to calculation at NLO.

The picture of the breaking of $R^{\gamma Z} = R^{\gamma}$ is qualitatively
similar for the deuteron, but with the size of the effect diminished by several
percent. As with the proton calculation, the strongest departure of the ratio
plotted in Figure \ref{fig:Fig_4} from unity is obtained at low $x$ --
in this case $x \approx 0.05 - 0.1$, at which $R^{\gamma Z} / R^{\gamma}
\approx 0.985$. This smaller effect is consistent with the nature of the
deuteron as an iso-scalar target, which leads to the large-scale cancellation
of flavor-dependence.
\begin{figure}[t]
\includegraphics[height=7.5cm,angle=270]{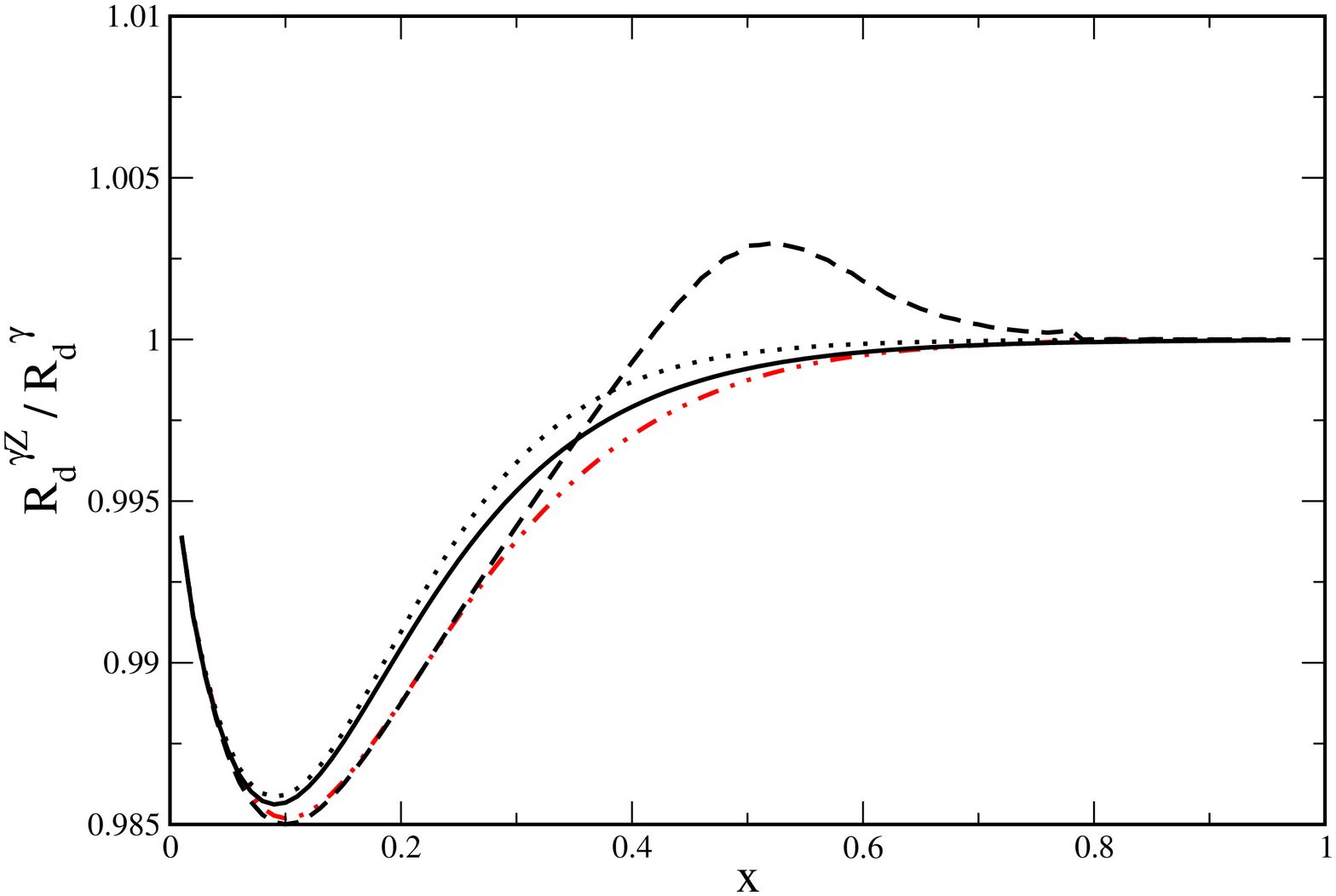}
\includegraphics[height=7.5cm,angle=270]{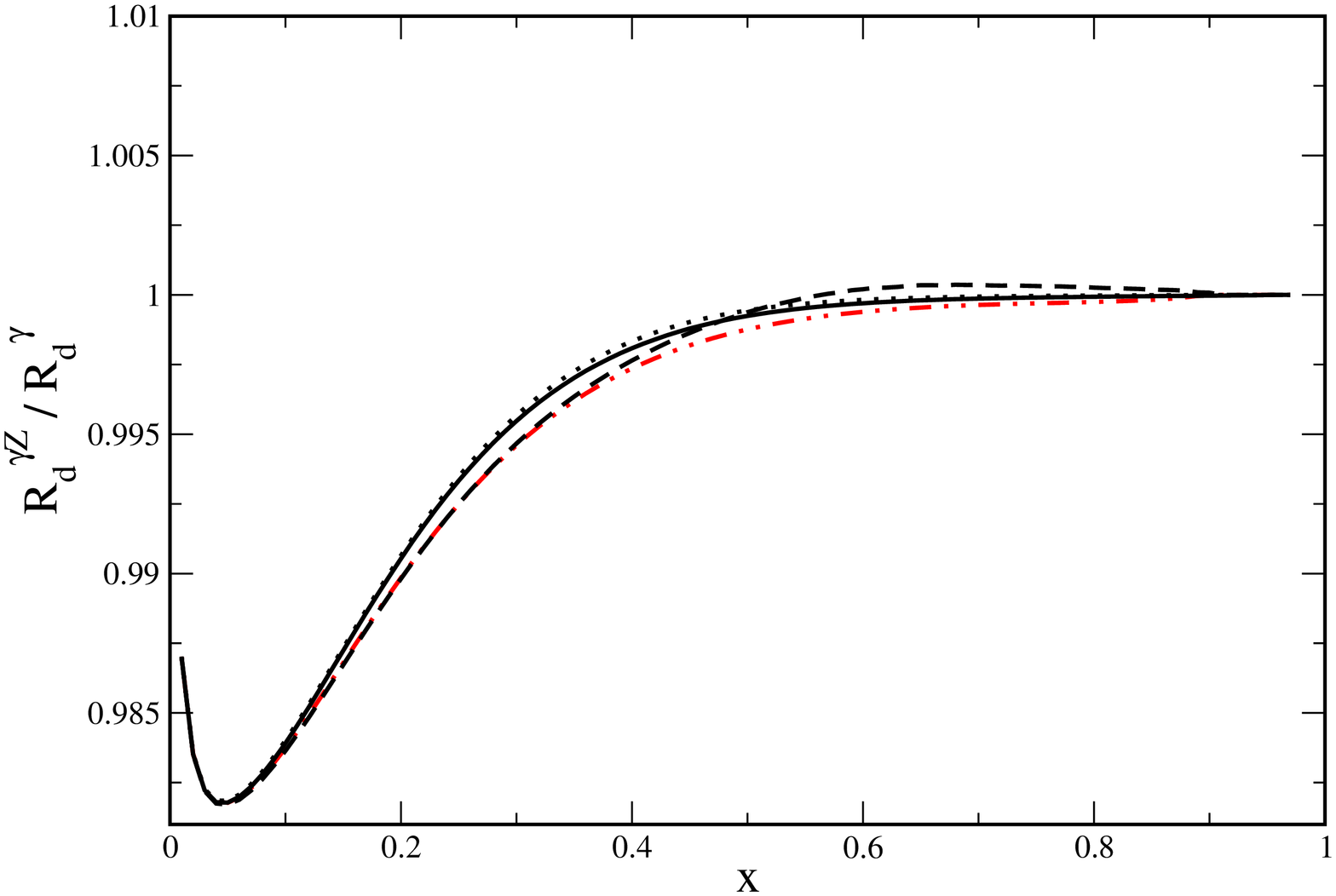}
\caption{Similar to Figure \ref{fig:Fig_3}, but for $R^{\gamma Z} / R^{\gamma}$
as determined in electron-deuteron scattering. Here, the nature of the deuteron
as an iso-scalar target softens the breaking of $R^{\gamma Z} = R^{\gamma}$. The
conventions for the linestyles are identical to Figure \ref{fig:Fig_3}.}
\label{fig:Fig_4}
\end{figure}

\begin{figure}[t]
\includegraphics[height=7.5cm,angle=270]{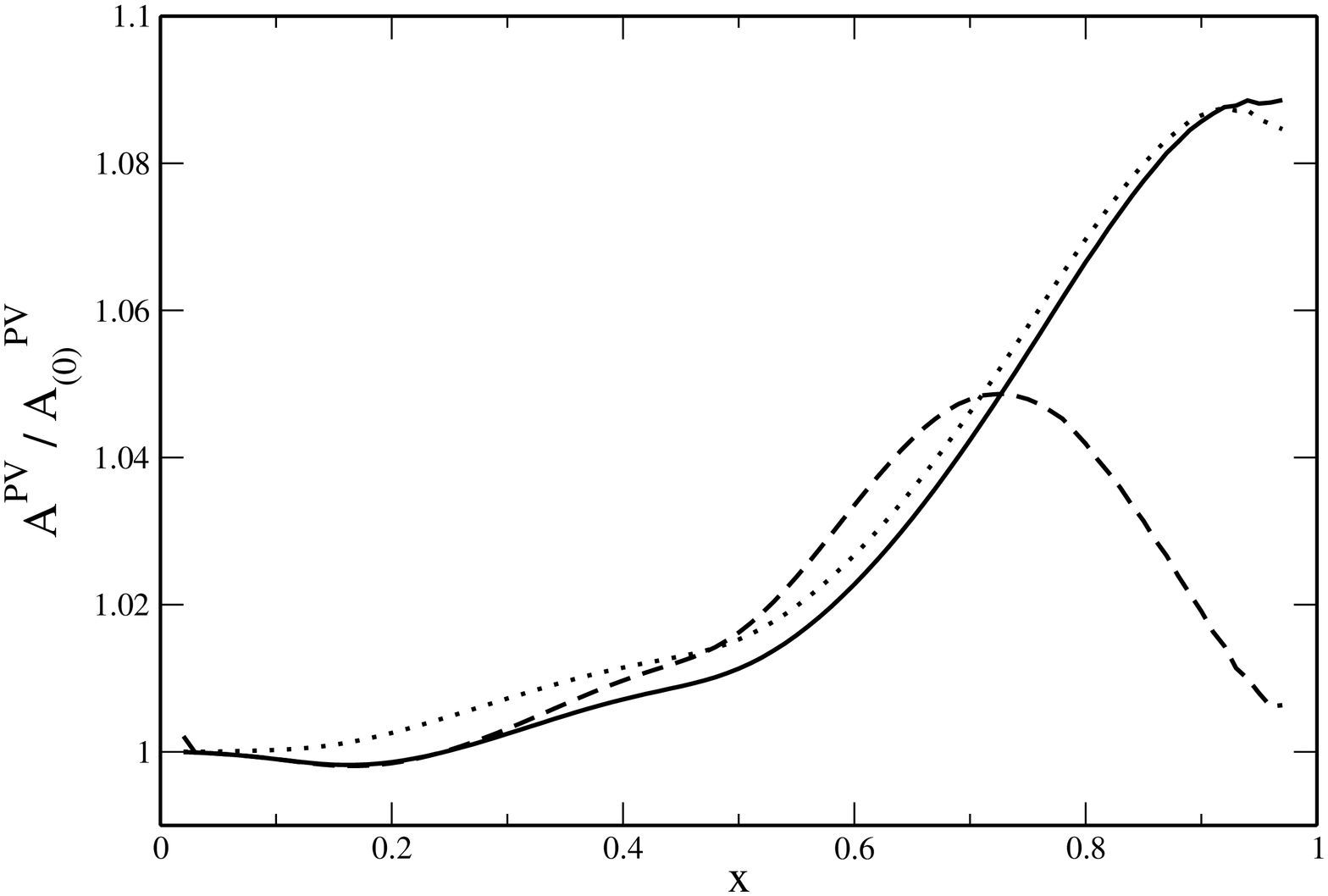}
\includegraphics[height=7.5cm,angle=270]{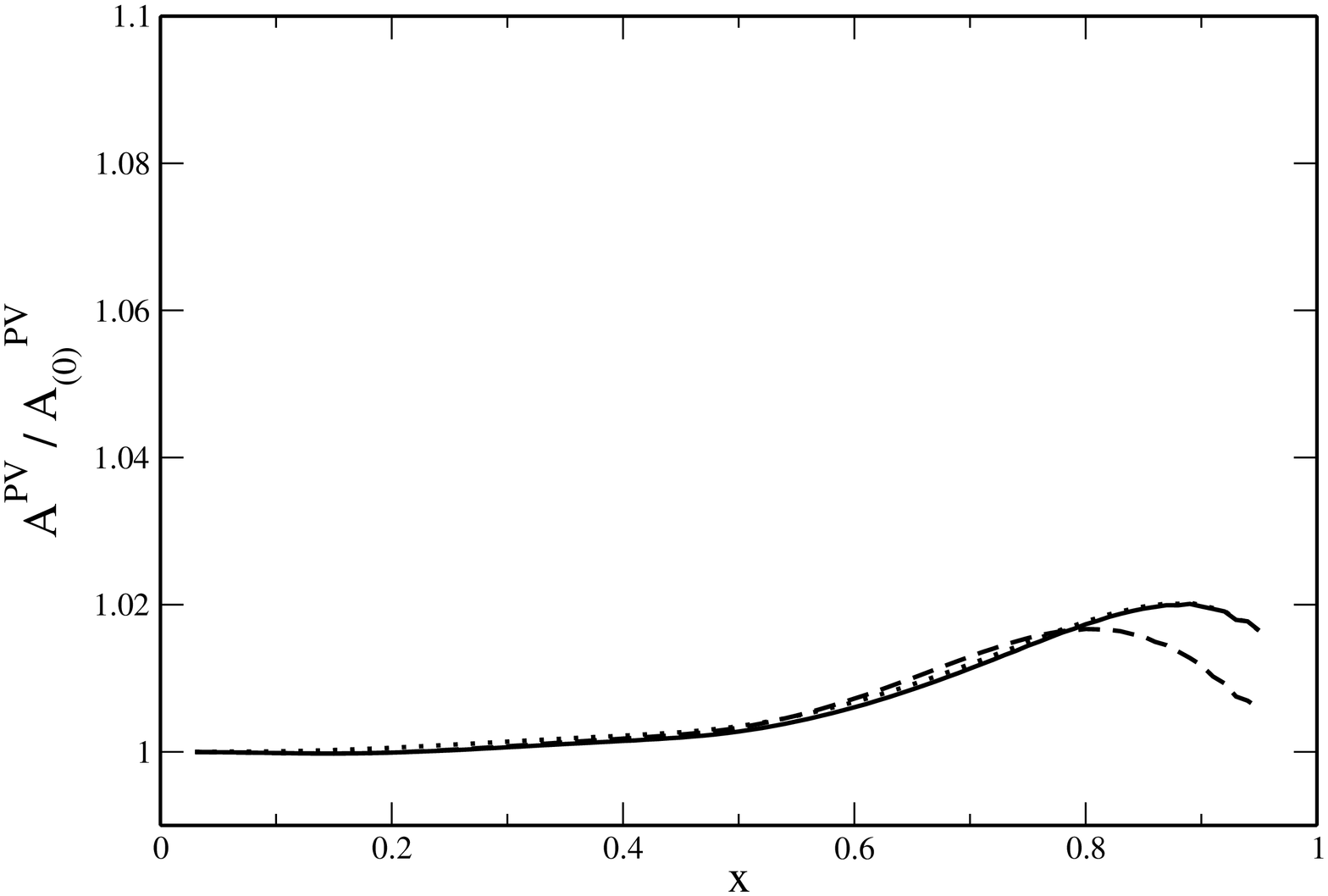}
\caption{The mass effect in the parity-violating asymmetry $A^{TMC}/A^{(0)}$ at
low and intermediate $Q^2$. We plot the mass effect in our three primary
prescriptions -- the OPE (solid), O(1/$Q^2$) expansion (dashed), and CF (dotted)
treatments.}
\label{fig:Fig_5}
\end{figure}

With some understanding of the behavior of $R^{\gamma Z}$ and $R^{\gamma}$ under
TMCs, we wish to perform a similar calculation in the full, parity-violating
asymmetry $A^{PV}$. As before, we consider a ratio of corrected/uncorrected
asymmetries in the various prescriptions; the result of this calculation at
low and intermediate $Q^2$ is given in Figure \ref{fig:Fig_5}.
For low $Q^2$, we note a relative similarity among the various 
prescriptions to intermediate $x \approx 0.6$ at which the effect of TMCs
is $\approx 6\%$; as one moves to largest $x$, the OPE and CF treatments
obtain the greatest effect ($\approx 8\%$), whereas the O(1/$Q^2$) expansion
of the LT OPE falls to zero. These effects are quickly suppressed as one
evolves to intermediate $Q^2 = 10$ GeV$^2$ for which the TMCs are
$\leq 2\%$, and the O(1/$Q^2$) expansion agrees with the other prescriptions
to higher $x$. This observation is consistent with the general message
for experimental efforts: though mass effects can be sizable
for lower values of $Q^2$, they may be brought into considerable control
by moving to modestly larger $Q^2$ at which the computed TMCs exhibit
less model-dependence.

It is natural to extend this calculation to the deuteron, for which the
property of iso-scalarity diminishes flavor-dependence as previously
noted; the resulting asymmetry is dependent only on the electroweak
couplings and the kinematical parameters $Y_1, Y_3$ \cite{Hobbs:2008mm}:
\begin{equation}
A^{\rm PV}
= -\left( \frac{3 G_F Q^2}{10 \sqrt{2}\pi\alpha} \right)
  \left[ Y_1 \left( 2 C_{1u} - C_{1d} \right)\
      +\ Y_3 \left( 2 C_{2u} - C_{2d} \right)\
  \right]\ ,
\end{equation}
where the $C_{1u}, C_{1d}$, etc are coupling constants. The disappearance
of the explicit dependence of the deuteron asymmetry on SFs leads to a very
small sensitivity to TMCs: generally, even at small $Q^2$, the mass effect in
the deuteron is sub-percent and model-independent in the sense that the
various prescriptions outlined here yield similarly small corrections. This
is promising for experimental efforts that aim (for instance) to precisely
extract the electroweak coupling constants from electron-deuteron scattering
events.
\begin{figure}[t]
\includegraphics[height=7.5cm,angle=270]{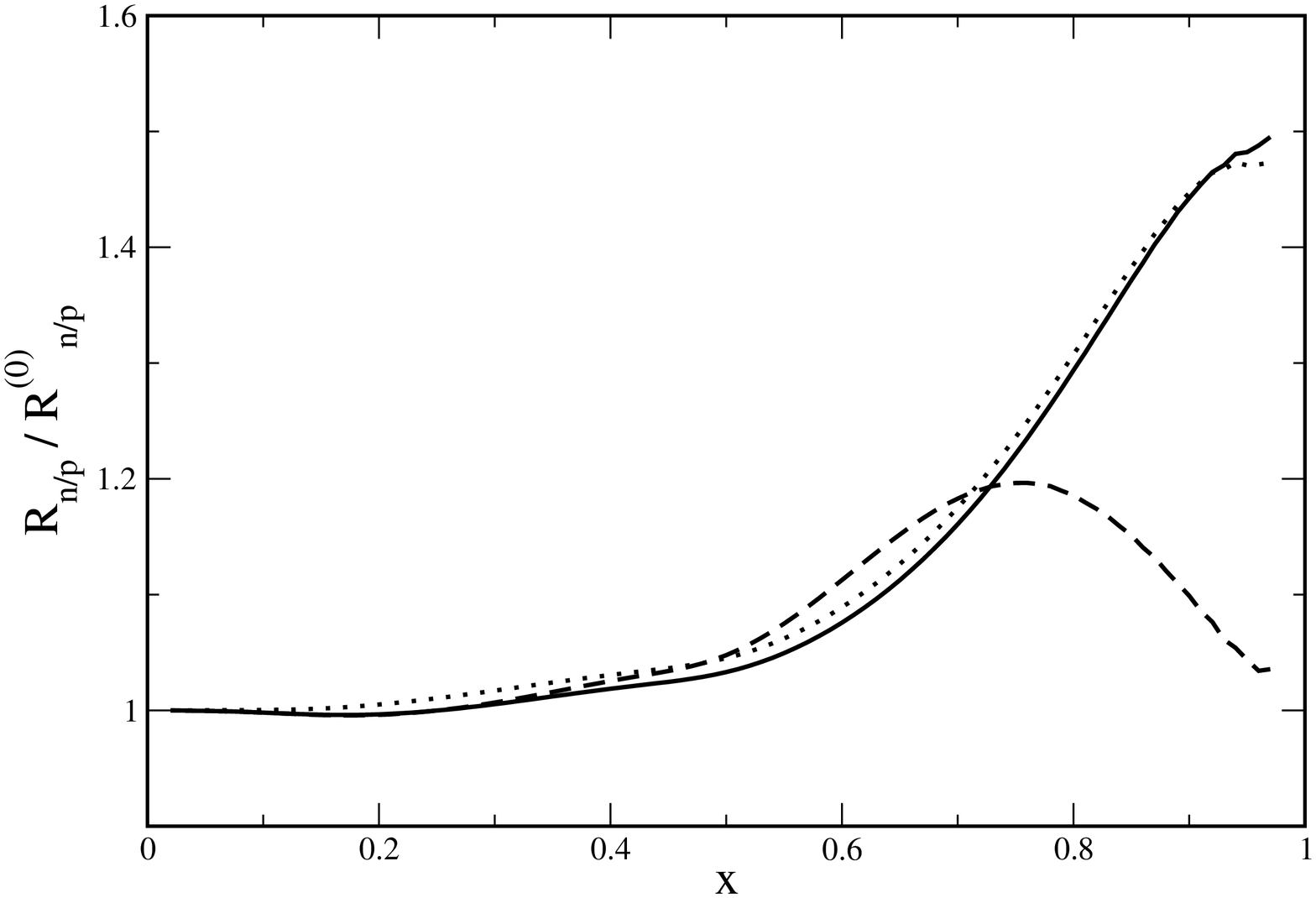}
\includegraphics[height=7.5cm,angle=270]{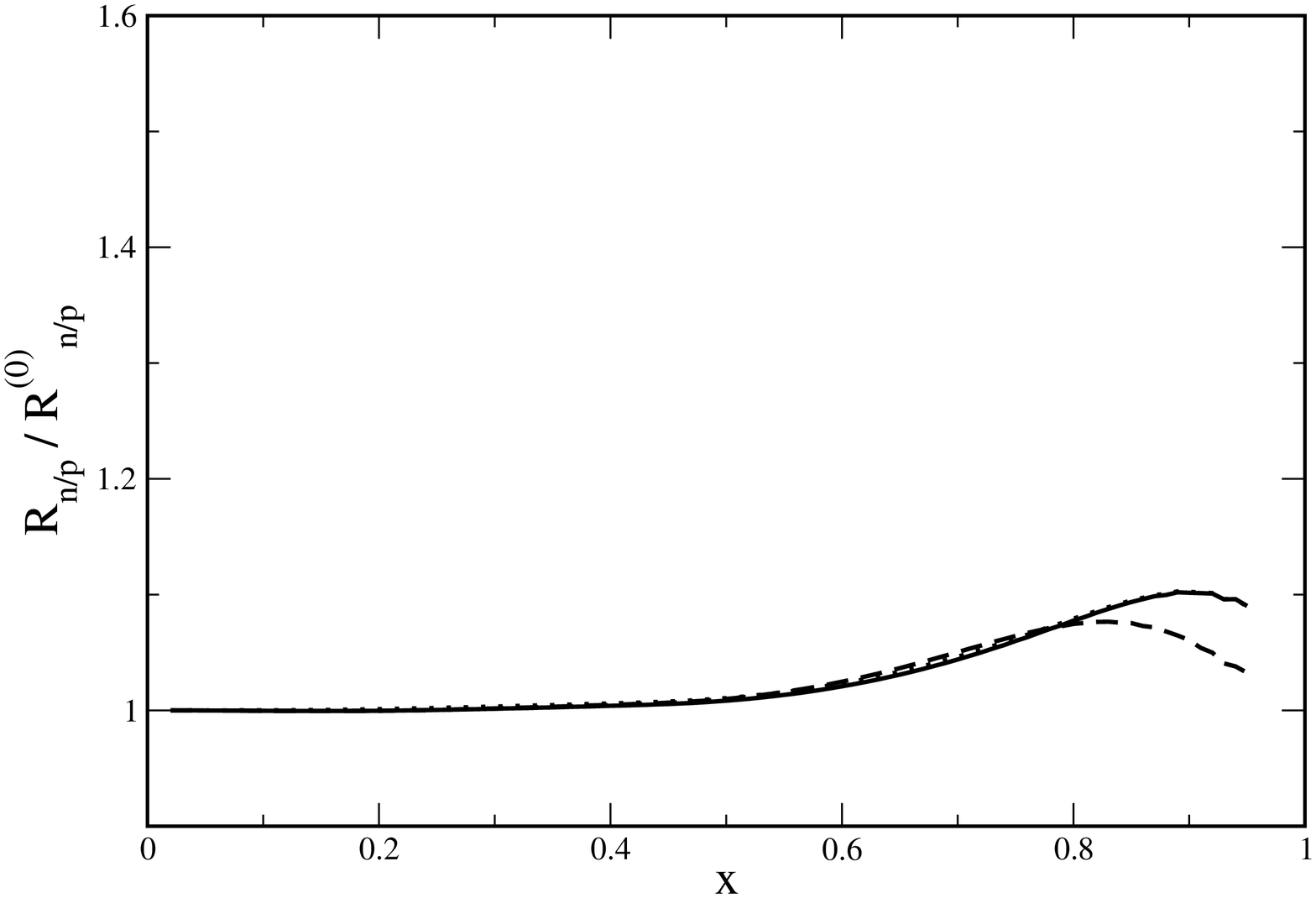}
\caption{The TMC effect in the SF ratio $R_{n/p} = F_2^n / F_2^p$. As with
Figure \ref{fig:Fig_5}, the prescriptions are the OPE (solid), O($1/Q^2$)
expansion (dashed), and CF (dotted).}
\label{fig:Fig_6}
\end{figure}

Lastly, as the SF ratio $R_{n/p} = F_2^n / F_2^p$ is
sensitive to the behavior of the PDF ratio $d/u$, it has attracted substantial
interest as a means of constraining various quark models. This is apparent if
we observe, for example, that if the large-$x$ ratio $d/u \rightarrow 0$, then
the SF ratio behaves as $R_{n/p} \rightarrow -\frac{1}{2} C_{1d} / C_{1u}
\approx 0.9$. Given this interest in the phenomenology of $R^{n/p}$ (particularly
at large $x$ where TMCs are often more pronounced), a picture of the mass
corrections to $R^{n/p}$ would be helpful. Such an illustration of the relative
mass effect is given in Figure \ref{fig:Fig_6}. As with $A^{PV}$ we plot a
correction ratio $R_{n/p} / R^{(0)}_{n/p}$ for $Q^2 = 2 (10)$ GeV$^2$, and obtain
a result which qualitatively closely resembles the full proton asymmetry -- up to
the scale of the effect. Here again, the size of the correction, as well as the
agreement among the various prescriptions are dramatically improved by
modestly evolving upward in $Q^2$.

In conclusion, while the TMCs and perturbative corrections in $\alpha_S$ are
considerable at lowest $Q^2$ where other $1/Q^2$ and log($Q^2$) effects become
important, they may be controlled in experimental efforts. In particular, performing
measurements at higher $Q^2$ and choosing iso-scalar targets enables one to cleanly
probe nucleon structure with less need to worry about possible model-dependent
mass effects. Still, the issues of the kinds of physics that might produce $R^{\gamma Z}
\neq R^{\gamma}$ and the possible implications for high-$x$ phenomenology remain
rich and largely unexplored topics.

\begin{acknowledgments}
We thank W. Melnitchouk, A. Accardi, J. Owens, M. Ramsey-Musolf,
K. Kumar, and P. Souder for valuable discussions and communications.
\end{acknowledgments}


\begin{thebibliography}{99}

\bibitem{Souder:2005tz}
  P.~A.~Souder,
  AIP Conf.\ Proc.\  {\bf 747}, 199-204 (2005).

\bibitem{Blumlein:1998nv}
 J.~Blumlein and A.~Tkabladze,
 Nucl.\ Phys.\  B {\bf 553} (1999) 427
 [arXiv:hep-ph/9812478].

\bibitem{Georgi:1976ve}
  H.~Georgi, H.~D.~Politzer,
  Phys.\ Rev.\  {\bf D14}, 1829 (1976).

\bibitem{Nachtmann:1973mr}
  O.~Nachtmann,
  Nucl.\ Phys.\  {\bf B63}, 237-247 (1973).

\bibitem{Schienbein:2007gr}
  I.~Schienbein, V.~A.~Radescu, G.~P.~Zeller {\it et al.},
  J.\ Phys.\ G {\bf G35}, 053101 (2008).
  [arXiv:0709.1775 [hep-ph]].

\bibitem{Kulagin:2004ie}
  S.~A.~Kulagin, R.~Petti,
  Nucl.\ Phys.\  {\bf A765}, 126-187 (2006).
  [hep-ph/0412425].

\bibitem{Accardi:2008ne}
  A.~Accardi, J.~-W.~Qiu,
  JHEP {\bf 0807}, 090 (2008).
  [arXiv:0805.1496 [hep-ph]].

\bibitem{Accardi:2009md}
  A.~Accardi, T.~Hobbs, W.~Melnitchouk,
  JHEP {\bf 0911}, 084 (2009).
  [arXiv:0907.2395 [hep-ph]].

\bibitem{Hobbs:2008mm}
  T.~Hobbs, W.~Melnitchouk,
  Phys.\ Rev.\  {\bf D77}, 114023 (2008).
  [arXiv:0801.4791 [hep-ph]].

\bibitem{Kulagin:2007ju}
  S.~A.~Kulagin, R.~Petti,
  Phys.\ Rev.\  {\bf D76}, 094023 (2007).
  [hep-ph/0703033 [HEP-PH]].

\bibitem{Whitlow:1990gk}
  L.~W.~Whitlow, S.~Rock, A.~Bodek {\it et al.},
  Phys.\ Lett.\  {\bf B250}, 193-198 (1990).
  
\end{thebibliography}
\end{document}